# STUDY OF THE PECULIAR VARIABLE STAR KHAVCHENKO 1

*A. E. Maslak, D. V. Denisenko\**

State education center "Vorobyovy gory" (Moscow Palace of Pioneers), Moscow, Russia

The work is devoted to the study of the peculiar variable star Khavchenko 1 in Auriga. It was discovered by the 7[th] grade student Sergey Khavchenko in 2020. The brightness of the star varies with an amplitude of about 0.4 magnitude and the period of 4.213 days. Initially the star could not have been attributed to any known type of variability. Its light curve looks like the upside-down light curve of the eclipsing binary similar to that of luminous red nova V1309 Scorpii a few years before the eruption. The analysis of the new data from the ZTF project gathered from 2020 till 2024 has shown the cyclic variations of the light curve amplitude. We have proposed several possible models of this system which can explain such an unusual behavior of this star. The continued observations of this star are necessary to reveal the period changes and to estimate the possible eruption date.

## Discovery

The variable star Khavchenko 1 was discovered by the 7[th] grade student Sergey Khavchenko in 2020 during the astronomical project camp at the "Komanda" Education center of Moscow Palace of Pioneers. The discovery was made on February 21[st] using the images of sky area around the cataclysmic variable DDE 180 obtained with the Autonomous Robotic Telescope (ART) at telescope.org. The MuniWin (C-Munipack) software was used to search for the variability on the images, and the period was determined using WinEffect program by V. P. Goranskij.

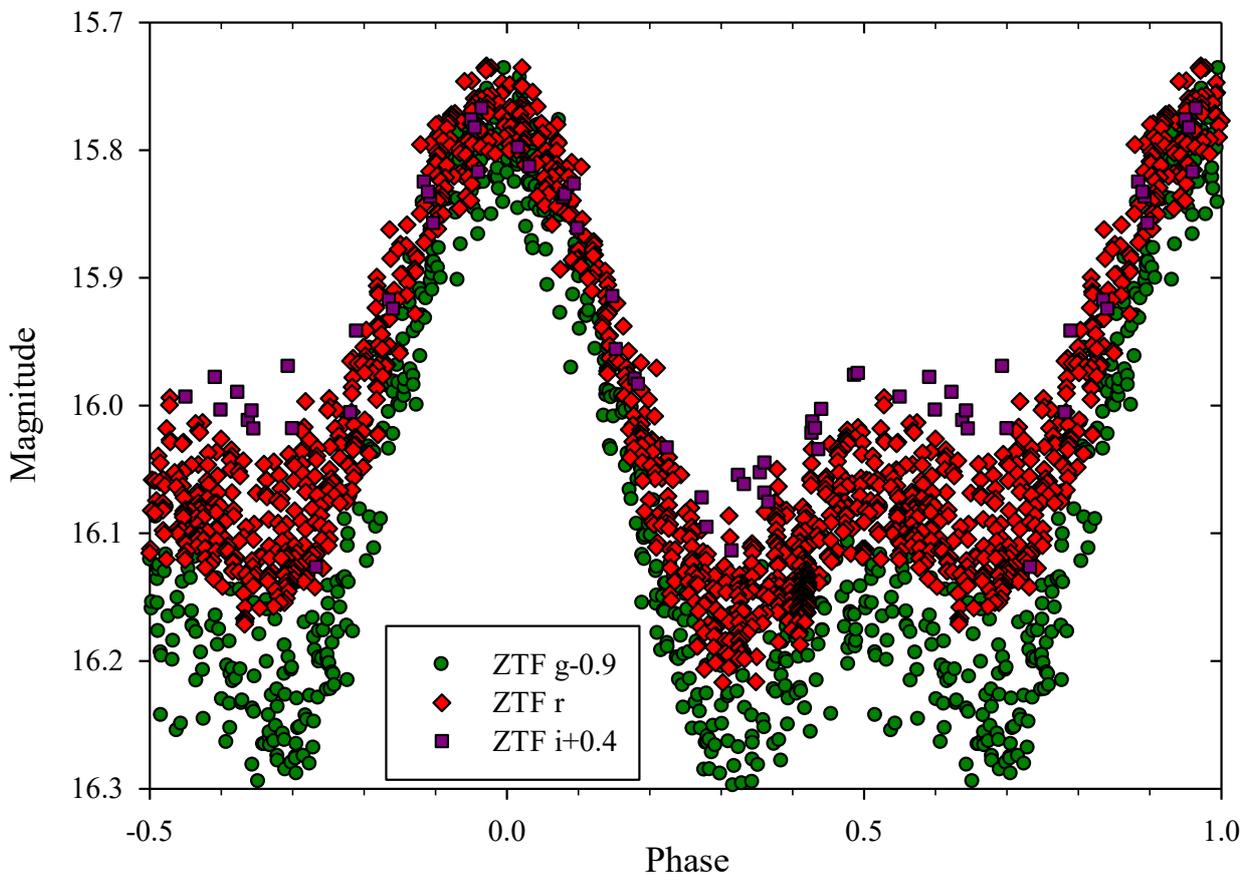

**Fig. 1.** Phase plot of Khavchenko 1 from ZTF data with the 4.2126 d period

Only 1.5 years of observations were available at the time of discovery. Based on this set of data a period of 4.2133±0.0008 days was determined. The star was submitted to the International Variable Star Index AAVSO VSX (Watson et al., 2006) with the name Khavchenko 1 as the variable of unknown type. Stars like this are labeled with an asterisk (*) in VSX with the following comment:

«Unique variable stars outside the range of the classifications. These probably represent either short stages of transition from one variability type to another or the earliest and latest evolutionary stages of these types, or they are insufficiently studied members of future new types of variables.»

Among 2.5 million objects in VSX database as of February 2020, only 18 stars (less than 0.001%) had the variability type *. Objects like those apparently deserve attention and require the following study.

**Improving the period of variability**

The 23$^{rd}$ data release of Zwicky Transient Facility (Masci et al., 2019) ZTF DR23 was made publicly available in January 2025. The photometric data for Khavchenko 1 is now covering 6 years of observations from 2018 to 2024 with more than 600 measurements in *g* and more than 1000 in *r*. WinEffect program was used for improving the period value. It has turned out to be 4.2126±0.0001 days (Fig. 1). The parameters of the new variable star are presented in Table 1. The moment of the main maximum is given as an initial epoch since it is the most outstanding light curve feature and can be more precisely determined rather than the minima.

Table 1. Parameters of Khavchenko 1

| Position (Epoch 2000.0) | 05 56 28.43 +32 38 56.8 |
|---|---|
| Constellation | Auriga |
| Magnitude range | 15.76–16.18 *r*, 16.70–17.18 *g* |
| Orbital period | 4.2126 days (101.1 hours) |
| Epoch of maximum light | JD=2458350.895 |
| Variability type | * |

**Study of the star**

The uniqueness of Khavchenko 1 is in the unusual shape of its phased light curve. Initially it was not even possible to attribute it to one of the major classes of variability (pulsating, eclipsing or rotating). All three classifications face the contradiction. As seen in Fig. 1, the light curve has two maxima of distinctly different heights and two minima of the same depth. Yet the phases of maxima are separated by 0.5 (half period), while the minima are not. Among the pulsating variables there are stars with the secondary hump which is lower than the main maximum, but the stars like those have remarkably asymmetric profile of the main peak (the fast rise and the slow decay). Besides, the secondary minimum in such stars (of RR Lyrae type) is strongly shifted from the mid-cycle and is usually placed at the phases of 0.75–0.8. The eclipsing nature of the variability appears to be more plausible, but even in this case the star is rather an exception. All but a few eclipsing binaries have the matching maxima and different minima, since the views at these stars on phases 0.25 and 0.75 (in profile from left and right) are no different. There are exotic binary systems with O'Connell effect, but even it cannot explain the difference between maxima of 0.4 magnitudes along with the same depth of minima as in Khavchenko 1. Finally, the version of rotating variability also has a right to exist, but it requires the presence of two giant cold spots with different temperatures and of different areas at the opposite sides of the single star which is exceptionally improbable.

The light curve appears remarkably different in various bands of visible light. While both minima are matching in *g* band, in *r* filter the secondary minimum is approximately 0.05 magnitude higher than the primary one, and in *i* filter the difference is getting as large as 0.1$^m$. It should be noted here that the minima fall on the phases of period differing from 0.25 and 0.75 by approximately 0.06 (0.31 and 0.69, correspondingly).

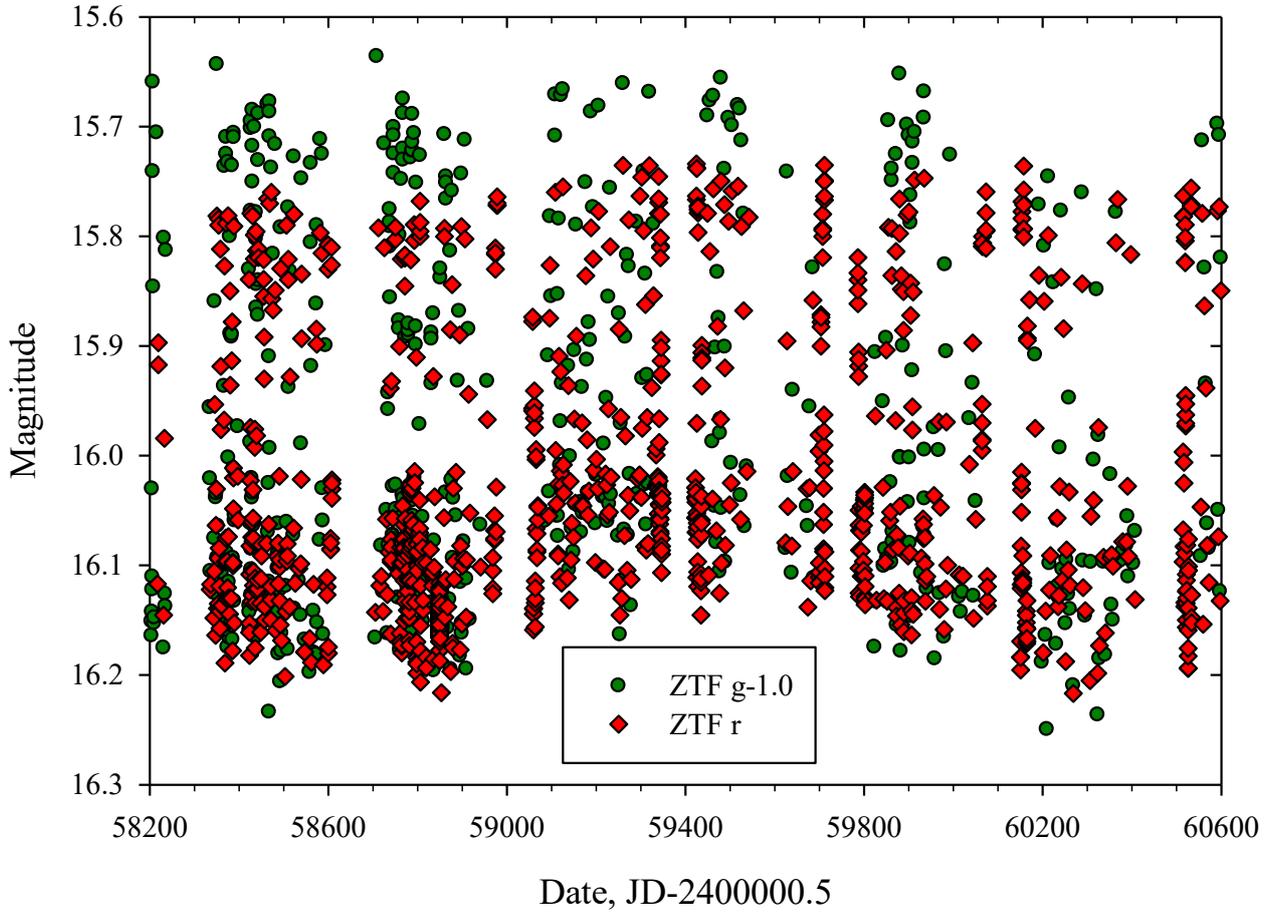

**Fig. 2.** Light curve of the variable star Khavchenko 1 from ZTF data. Green circles are observations in *g* filter, red diamonds – in *r* filter. Measurements in *g* band are raised by 1 magnitude.

It is obvious from the 6-years ZTF light curve that the maxima and minima are varying by about 0.1 magnitude. The search for period in the range from 1200 to 1800 days has revealed the best value of 1550±60 days (4.2±0.2 years).

Using the 3$^{rd}$ Data Release of Gaia space observatory (Gaia collaboration, 2022) one can convert the parallax value of 0.559±0.056 mas to the distance of Khavchenko 1 as 1800±180 parsecs. The absolute magnitude in the Gaia photometric band $M_G$ is then +4.7. The star belongs to the spectral type G or K, thus making the pulsating nature of variability improbable. Sun-like stars do not show the pulsations with the amplitudes of 0.4$^m$ and the periods of 4–5 days.

Taking all the above into account we have settled on the scenario of eclipsing binary star with peculiar characteristics. Three possible models of Khavchenko 1 are proposed to explain the unusual light curve shape of this star.

**Possible physical models**

Fig. 3 shows the model of Khavchenko 1 with the cold spots, Fig. 4 – with the hot spot on one of the components, and finally Fig. 5 – the model of interacting binary system with the accretion stream from one component onto the other.

In principle, all three models above can explain the unusual behavior of Khavchenko 1. Yet every one of them has its own advantages and drawbacks. Let us consider these models one by one to select the most non-contradictory one that agrees better with the observational data.

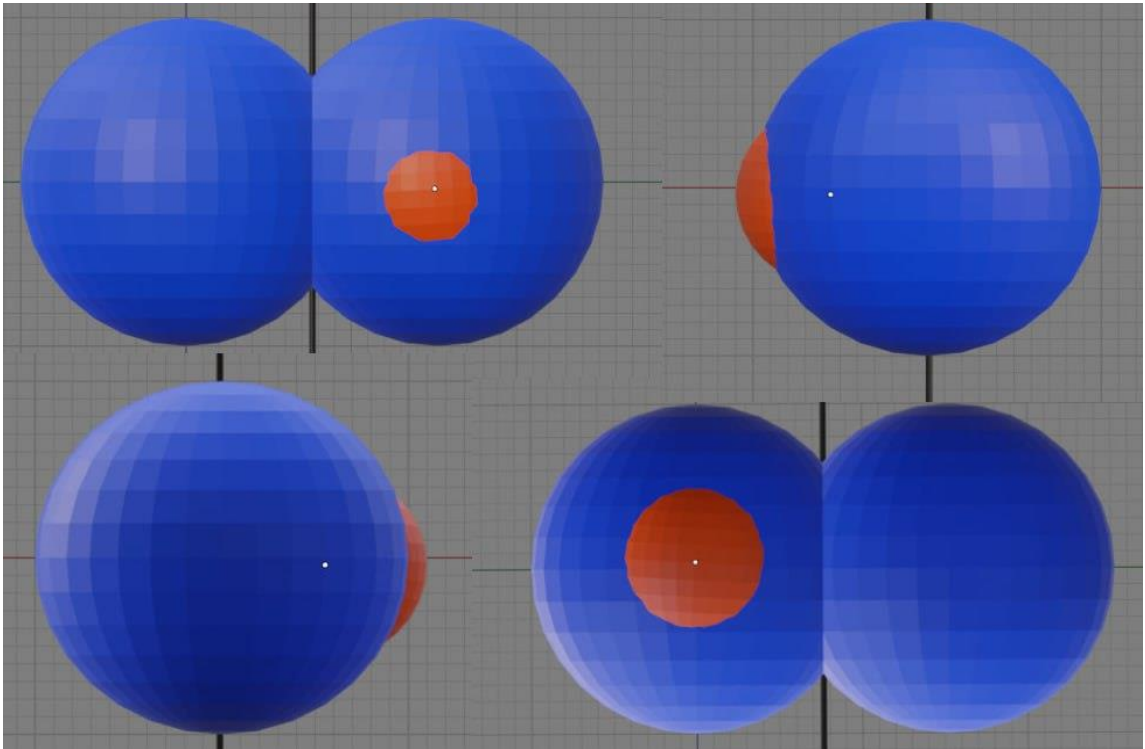

**Fig. 3.** Model of the binary star Khavchenko 1 with the cold spots

Model with the cold spots explains well the cyclic variations of maxima and minima if one supposes 4.2 years to be the activity cycle similar to the 11-year Solar one. The activity of the Sun is known not to be strictly periodic, with the maxima and minima occurring as early as 8 years and up to 18 years after the previous one and having different heights. The ZTF data is covering only one cycle of Khavchenko 1 activity and a half which does not allow us to draw conclusions about the stability of the 4-years cycle. The main objection against this model can be the very large difference between the primary and secondary maxima which requires the presence of a very cold and huge spot on one side of the star covering about 40% of its surface.

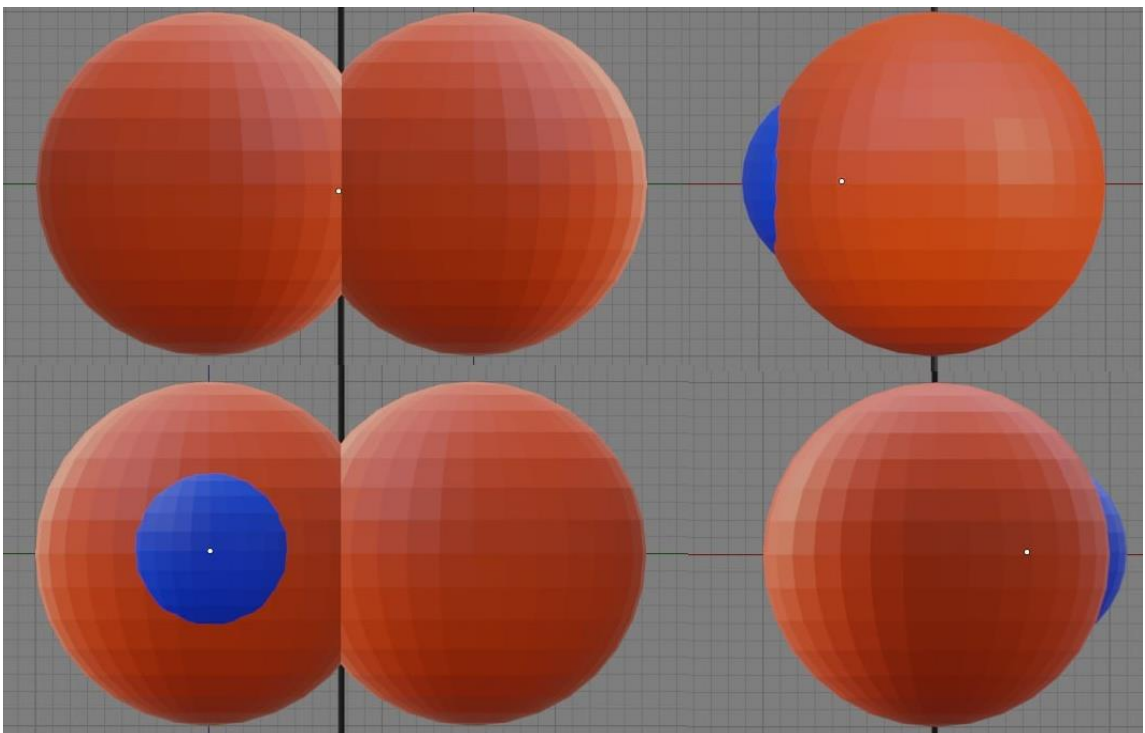

**Fig. 4.** Model of the binary star Khavchenko 1 with the hot spot

Model with the hot spot appears to be less likely since it is hard to imagine the mechanism causing the cyclic variations of the average star brightness. Moreover, this model does not explain the asymmetric shape of the minima and especially the rise of the secondary minimum above the primary one in the red light (*r* and *i* filters).

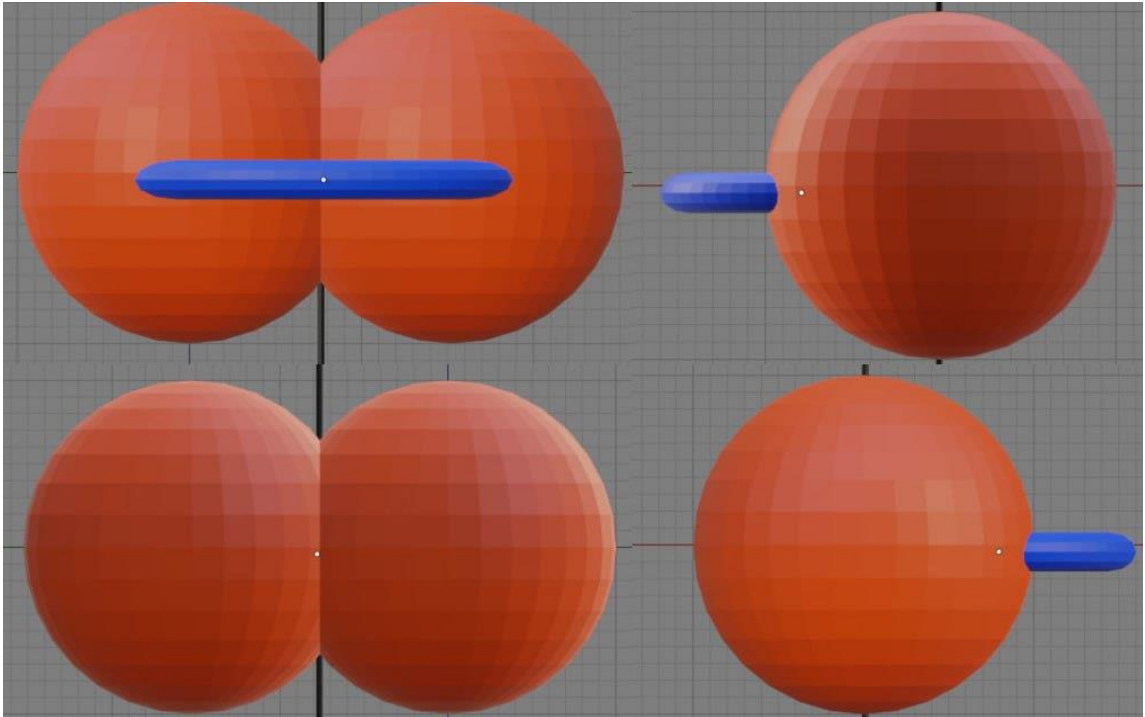

**Fig. 5.** Model of the binary star Khavchenko 1 with the accretion stream

Finally, the model with matter streaming from one component to the other is the most promising one. In this model the binary system configuration differs significantly from the symmetric one which explains the different magnitude at the phases around 0.25 and 0.75. In case of two spherical components, as already told above, the light curve should remain the same after the phase reversal. As to the variation of minima and maxima with the cycle of about 4 years, it can be caused by the orbital plane precession, as well as by the non-stationary accretion rate due to the magnetic activity of one or both components of the system.

**Search for possible analogs**

Among the objects labeled as * in VSX database we found only one more star with the light curve similar to that of Khavchenko 1. This is the variable BMAM-V397 discovered by Mariusz Bajer in 2019. It has a period of 2.7091 days, amplitude of variations $1.4^m$ (18.2–19.6 *g*) and displays an even more extreme difference between the primary and secondary minima ($0.7^m$ in *r* filter and $0.9^m$ in *g* filter). BMAM-V397 has the color index (*g-r*) about 1.1 like that of Khavchenko 1 and the similar absolute magnitude $M_G = +4.3$.

But of most interest among the known variable stars from the similarity to Khavchenko 1 point of view is the luminous red nova V1309 Scorpii which has shown an outburst by 12.5 magnitudes in 2008. This star has turned out to be the result of the merger of two usual stars and was a contact binary system with a period of about 1.44 days before the eruption. During five years from 2002 to 2007 the orbital period of V1309 Sco has reduced from 1.44 down to 1.42 days, which was only found in the archival data post factum. But the most interesting thing is that the phase plot of V1309 Sco before merging was strongly reminding the light curve of Khavchenko 1, showing the maxima of different heights and the minima of the same depth (Tylenda et al., 2011).

## Conclusion

The study of the peculiar variable star Khavchenko 1 has shown that it is the contact binary system similar to the red luminous nova V1309 Sco before merging. The cyclic changes of the light curve amplitude are found with the period about 4.2 years. Three models of this system that we proposed could explain the observed behavior of the star. The first model implies cyclic activity of the cold spots similar to the solar ones, the second one – the presence of the hot spot on the surface of one component, and the third one – the accretion stream between the components and the possible orbital plane precession. The period of Khavchenko 1 is currently about 3 times longer than that of V1309 Scorpii. The follow up observations of this star are necessary to reveal the period changes and to estimate the possible merging time.